\documentclass[aps,pre,english,showpacs,showkeys,preprintnumbers,twocolumn,floatfix,nofootinbib,10pt]{revtex4-2}
\usepackage{eurosym}
\usepackage{amssymb}
\usepackage{graphicx,color}
\usepackage{epsfig}
\usepackage{rotating}
\usepackage[T1]{fontenc}
\usepackage{latexsym,epsfig}
\usepackage[latin9]{inputenc}
\usepackage{amsfonts}
\usepackage{dcolumn}
\usepackage{bm}
\usepackage{babel}
\usepackage{hyperref}
\usepackage{amsmath,mathrsfs}

\begin{document}

\title{First numerical evidence of Janssen-Oerding's prediction in a
three-dimensional spin model far from equilibrium}
\author{Roberto da Silva}
\address{Instituto de F{\'i}sica, Universidade Federal do Rio Grande do Sul,
Av. Bento Gon{\c{c}}alves, 9500 - CEP 91501-970, Porto Alegre, Rio Grande do
Sul, Brazil}

\begin{abstract}
Jansen and Oerding [H. K. Janssen, K. Oerding, J. Phys. A: Math. Gen. 27,
715 (1994)] predicted an interesting anomalous tricritical dynamic behavior
in three-dimensional models via renormalization group theory. However, we
verify a lack of literature about the computational verification of this
universal behavior. Here, we used some tricks to capture the log corrections
and the parameters predicted by these authors using the three-dimensional
Blume-Capel model. In addition, we also performed a more detailed study of
the dynamic localization of the phase diagram via power laws optimization.
We quantify the crossover phenomena by computing the critical exponents near
the tricritical point.
\end{abstract}

\maketitle

%\pacs{05.10.-a; 02.70.Tt, 05.70.Ln}

Blume-Capel (BC) model \cite{Blume-Capel} is a spin-1 model whose
Hamiltonian is:

\begin{equation}
\mathcal{H}=-J\sum\limits_{\left\langle i,j\right\rangle }\sigma _{i}\sigma
_{j}+D\sum\limits_{i=1}^{N}\sigma _{i}^{2}-H\sum\limits_{i=1}^{N}\sigma _{j}%
\text{.}  \label{Eq:short_range_hamiltonian}
\end{equation}%
Here, $D\geq 0$ is the anisotropy term, $\sigma _{j}=0,\pm 1$, and $H$ is
the external field that couples with each spin and $\left\langle
i,j\right\rangle $ denotes that sum is taken only over the nearest neighbors
in a $d$-dimensional lattice.

Such a model in two and three dimensions presents a critical line and a
first-order transition phase line, and such lines have an intersection point
known as a tricritical point (TP). Such a name is because for $H>0$ and $H<0$%
, one has two other first-order lines in addition to one from $H=0$, and all
these three lines culminate in that point. If the equilibrium studies of
this model are fascinating, their dynamic aspects are even more, mainly when
studied at TP.

Janssen, Schaub, and Schmittmann \cite{Janssen1989} proposed a dynamic
scaling relation that includes the dependence on the initial trace of the
system. This approach predicts an initial anomalous slip of magnetization on
the relaxation of a spin model that, initially at high temperatures ($%
m_{0}<<1$), is suddenly placed at its critical temperature. A power law with
exponent $\theta $ $=(x_{0}-\beta /\nu )/z>0$ describes such behavior, which
depends on universal exponents: the dynamic one z and the static exponents $%
\beta $ and $\nu $, these last ones related to the equilibrium of the
system. An anomalous dimension $x_{0}$ related to initial magnetization
completes its dependence.

Zheng and many collaborators (for a review, see \cite{Zheng1998})
numerically explored such scaling relation via MC simulations under many
aspects. In the sequence, many other authors enriched the method by
proposing new amounts, refinements, and other models, including also that
ones without defined Hamiltonian, and even models with long-range
interaction (see, for example \cite%
{Albano,Agente2,Agente3,Agente4,Agente5,Agente6,Agente7}).

The consequences of this theory, at criticality, is resumed as a transition
between two power laws:

\begin{equation}
m(t)=\left\{ 
\begin{array}{lll}
m_{0}t^{\theta } & \text{for} & t_{0}<t<m_{0}^{-z/x_{0}} \\ 
&  &  \\ 
t^{-\lambda } & \text{for} & t>>m_{0}^{-z/x_{0}}%
\end{array}%
\right.  \label{Eq:power_law_1}
\end{equation}%
where $\lambda =z^{-1}\beta /\nu $ and $m(t)$ is the magnetization per spin.
One way to check the second tail $m(t)\sim t^{-\lambda }$of this behavior is
to prepare systems from a wholly ordered initial system ($m_{0}=1$). In the
two-dimensional Blume-Capel model, time-dependent MC (TDMC) simulations show
exactly such behavior of its critical points ($D\geq 0$). However, for the
TP, such simulations show that $\theta $ is negative as theoretically
predicted by Janssen and Oerding \cite{Janssen1994} and via time-dependent
Monte Carlo (MC) simulations by R. da Silva et al. \cite%
{My-Tricritical-contributions}. This previous work showed that the magnitude
of this exponent is more than double the ones found for the critical ones
(Ising-like points).

Grasberger \cite{Grassberger} and Jaster et al. \cite{Jaster} initially
studied the tridimensional kinetic spin-1 Ising model (Blume Capel for $D=0$%
) using TDMC simulations to obtain the exponents of the model for the
critical point of this model. However, what happens when $D>0$? The behavior
described by Eq. \ref{Eq:power_law_1} remains valid for critical and
tricritical points? Can TDMC simulations show the crossover effects between
critical line (CL) and tricritical point (TP)?

This paper will explore the critical behavior of the three-dimensional
Blume-Capel model compared with the results from its version in two
dimensions via TDMC. We will show solid numerical evidence of the
log-corrections for the TP in its three-dimensional version theoretically
predicted by Janssen and Oerding \cite{Janssen1994}. We complete our study
estimating critical and tricritical parameters with a refinement method of
the power laws. The computation of critical exponents along the critical
line captures the crossover effects at proximities of the tricritical point.

We start our study by computing the coefficient of determination that here
measures the "quality" of power-law \cite{SilvaPRE2012}:

\begin{equation}
r=\frac{\sum\limits_{t=t_{\min }}^{t_{\max }}(\overline{\ln m}-a-b\ln t)^{2}%
}{\sum\limits_{t=t_{\min }}^{t_{\max }}(\overline{\ln m}-\ln m(t))^{2}},
\label{determination_coefficient}
\end{equation}%
with $\overline{\ln m}=\frac{1}{(t_{\max }-t_{\min })}\sum\nolimits_{t=t_{%
\min }}^{t_{\max }}\ln m(t)$. After a previous study of size systems, one
used systems with linear dimension$\ L=40$ ($N=L^{3}=6.4\times 10^{4}$
spins). Here $m(t)=\frac{1}{N_{run}N}\sum_{j=1}^{N_{run}}\sum_{i=1}^{N}%
\sigma _{i,j}(t)$ with $\sigma _{i,j}(t)$ denotes the $i$-th spin state at $%
j $-th run, at time $t$. We obtained such amount by performing averages over 
$N_{run}=300$ different runs (time evolutions). We also used $t_{\min }=10$
and $t_{\max }=100$ MC steps for our estimates.

\begin{figure}[tbp]
\begin{center}
\includegraphics[width=1.0\columnwidth]{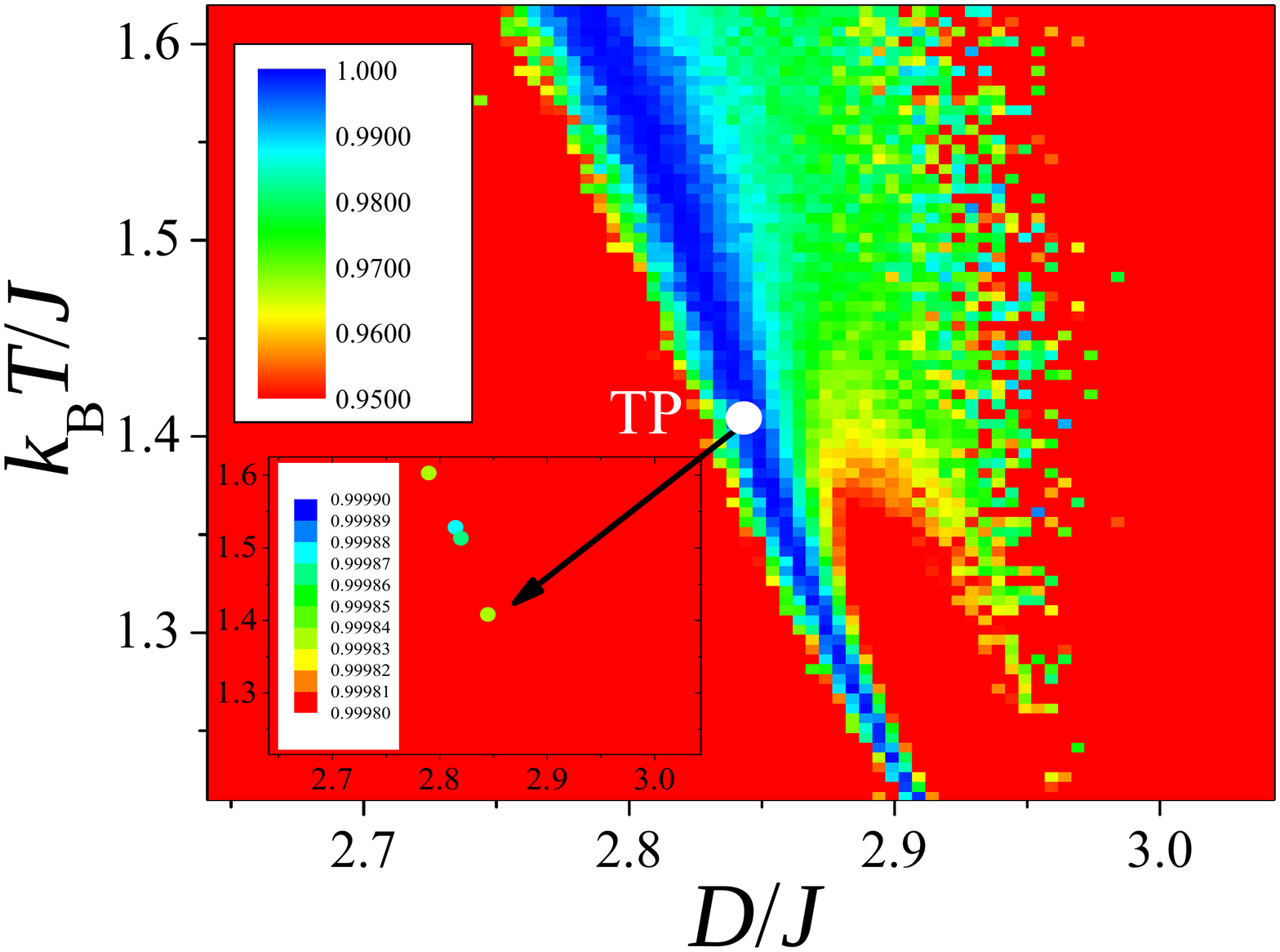}
\end{center}
\caption{Coefficient of determination for different parameters $k_{B}T/J$
and $D/J$. Until the point $k_{B}T/J\approx 1.418$ and $D/J\approx 2.845$
(TP according to literature estimates), the narrow blue region contains the
critical line. The inset plot shows the reminiscent points over a
significantly restricted situation: $0.9998<r<1$, showing that the
optimization does not find other points after the TP point in this
situation. \ \ \ \ \ \ \ \ \ \ \ \ \ \ \ \ \ \ \ \ \ \ \ \ \ \ \ \ \ \ \ \ \
\ \ \ \ \ \ \ \ \ \ \ \ \ \ \ \ \ \ \ \ \ \ \ \ \ \ \ \ \ \ \ \ \ \ \ \ \ \
\ \ }
\label{Fig:coeficient_of_determination_heat_map}
\end{figure}

We vary $k_{B}T/J$ from $1.218$ until $1.618$, from $D/J=2.645$ until $3.040$
with values spaced of $\Delta =2.5\times 10^{-3}$ for both parameters. This
diagram ( Fig. \ref{Fig:coeficient_of_determination_heat_map}) shows a
suggestive narrow region (blue) that includes the critical line since it
contains the points with the highest coefficients of determination, i.e.,
candidates to the critical points. The region becomes narrower as it
approaches the TP (see, for example, \cite{Deserno} $D/J=2.84479(30)\ $and $%
k_{B}T/J=1.4182(55)$) for the tricritical coupling ratio, which is a
\textquotedblleft foreshadowing\textquotedblright\ of the crossover effects.
After this point, it becomes even narrower, and this is only an
\textquotedblleft echo\textquotedblright\ of the critical region since one
expects only a first-order transition for $D/J$ $\geq 2.8502$ \cite{Butera}
and in this case, out of figure since this first-order transition point
corresponds to temperature: $k_{B}T/J=0.221(1)$.

In addition, it is essential to mention that if we perform a severe
restriction to the coefficient of determination: $0.9998<r<1$, one does not
observe points after $D/J>2.84479$\ (TP), as observed in the inset plot in
Fig. \ref{Fig:coeficient_of_determination_heat_map}. This corroborates the
fact that these extra blue points found in the original figure were, as
previously mentioned, only a \textquotedblleft
reverbaration\textquotedblright\ of the critical region and that the method
of coefficient of determination is reliable indeed.

Nevertheless, the optimal points are indeed the critical line points? By
using the critical points presented in Butera and Pernici \cite{Butera}
obtained via low and high-temperature expansions (see table 5 in this
reference), we can check if our critical points are precisely well estimated.

\begin{figure}[tbp]
\begin{center}
\includegraphics[width=1.0\columnwidth]{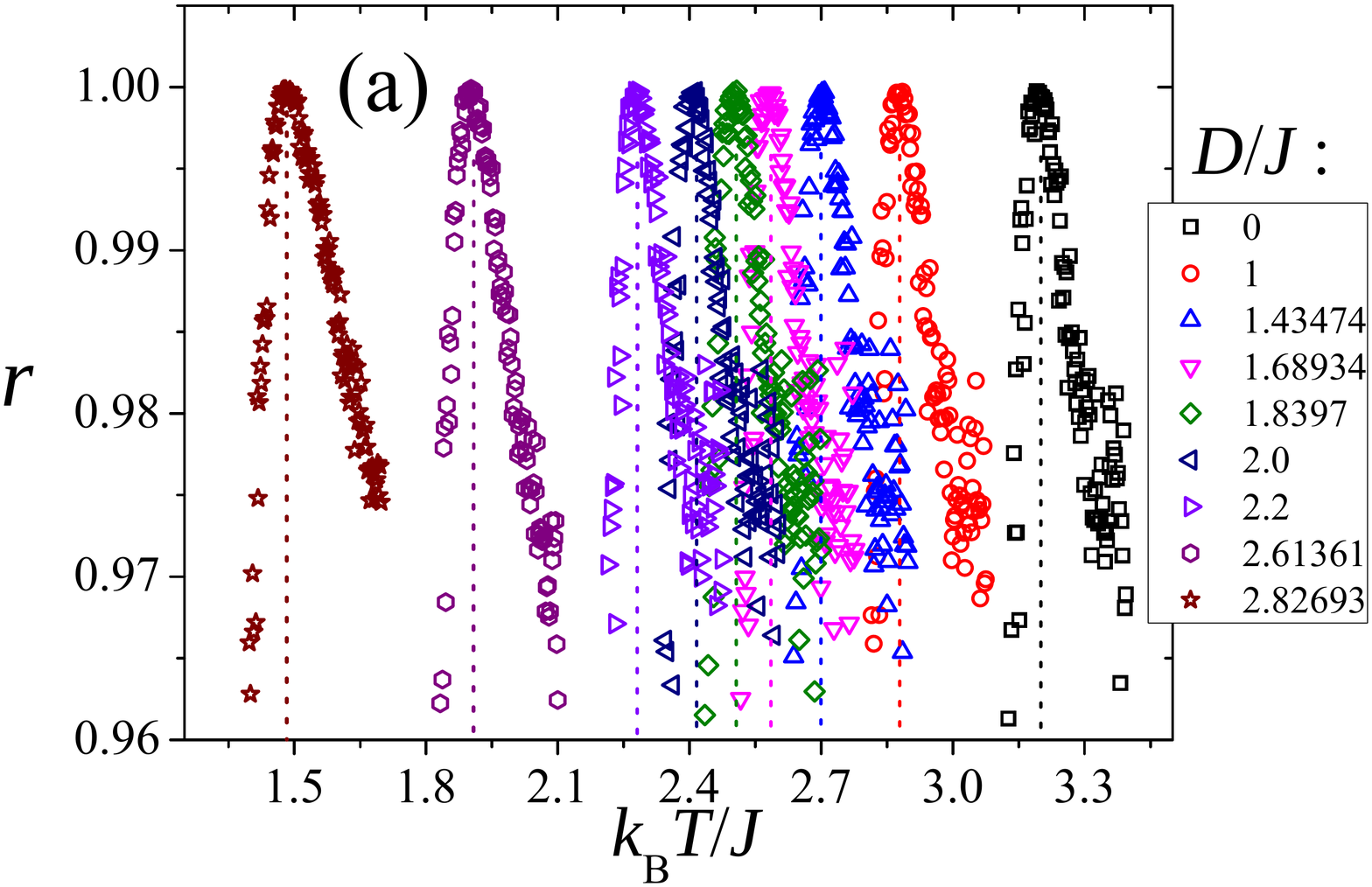} %
\includegraphics[width=1.0\columnwidth]{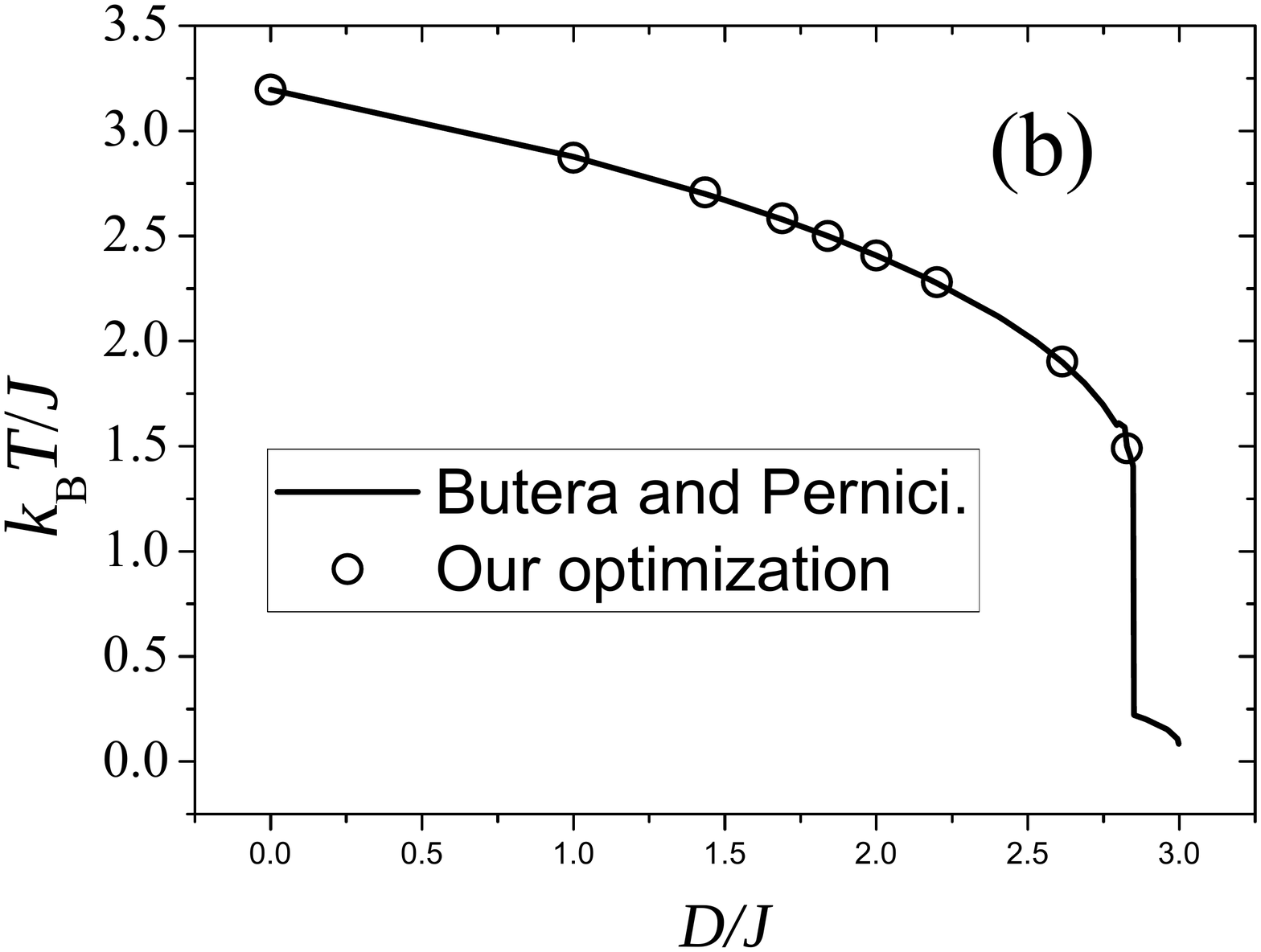}
\end{center}
\caption{(a) Curves $r$ $\times (k_{B}T)/J$ for some values of $D/J$. (b)
The points obtained with the optimization in (a) compared with the curve
obtained by Butera and Pernici \protect\cite{Butera}. }
\label{Fig:critical_line_optimization_comparison_Butera}
\end{figure}

We fixed some values of $D/J$ picked up from this same table. For each input 
$D/J$, we obtained the optimal corresponding value $k_{B}T/J$, which
corresponds to the maximal r value (see Fig. \ref%
{Fig:critical_line_optimization_comparison_Butera} (a) ). With these points
in hands, we compared with the critical line obtained by Butera and Pernici 
\cite{Butera} as described in Fig. \ref%
{Fig:critical_line_optimization_comparison_Butera} (b) whose used
equilibrium numerical methods. We observed an excellent match with such
results method, showing that we can obtain the critical values of the
three-dimensional Blume-Capel using time-dependent MC simulations with the
refinement method based on the coefficient of determination.\textbf{\ }

And about crossover effects? How is the sensitivity of these exponents as
they approach the tricritical point? For that, we look at different time
evolutions. First, to calculate the exponent $\theta $, we should study the
system with varying values of $m_{0}$ by performing an extrapolation $%
m_{0}\rightarrow 0$. Alternatively, we use a more accessible alternative
proposed by Tome and Oliveira \cite{TomeOliveira1998} by calculating: 
\begin{equation*}
C(t)=\frac{1}{N^{2}N_{run}}\sum_{i=1}^{N}\sum_{j=1}^{N_{run}}\sigma
_{i,j}(t)\sigma _{i,j}(0)\text{.}
\end{equation*}%
Such estimate considers $\sigma _{i,j}(0)$ randomly drawn ($0$, $-1$, or $+1$%
, with probability 1/3), such that $m(0)=\frac{1}{N_{run}N}%
\sum_{i=1}^{N}\sum_{j=1}^{N}\sigma _{i,j}(0)\approx 0$, which yields $%
C(t)\sim t^{\theta }$ when $N_{run}$ is large enough. For this experiment,
one used $N_{run}=30000$ runs, and we measured the slopes in the interval
[30,150] MC steps. The exponent $\lambda $ was obtained by performing
simulations starting with $m_{0}=1$ of $m(t)$. In this case, one used $%
N_{run}=300$ runs since simulations with $m_{0}=1$ require many fewer runs.
In order to obtain the exponent $z$, one simulates $m^{2}(t)=\frac{1}{%
N_{run}N^{2}}\sum_{j=1}^{N_{run}}\left( \sum_{i=1}^{N}\sigma
_{i,j}(t)\right) ^{2}$ by starting with $m_{0}=0$, and thus one considers
the ratio \cite{SilvaPLA2002} $F_{2}(t)=$ $\frac{m^{2}(t)_{m_{0}=0}}{\left(
m(t)_{m_{0}=1}\right) ^{2}}$, which behaves as $F_{2}(t)\sim t^{d/z}$. For $%
m^{2}(t)_{m_{0}=0}$, only $N_{run}=1000$ runs are enough for good estimates.
For estimates of $\lambda $ and $z$, we performed fits in the interval
[10,100] MC steps.

\begin{figure}[tbp]
\begin{center}
\includegraphics[width=1.0\columnwidth]{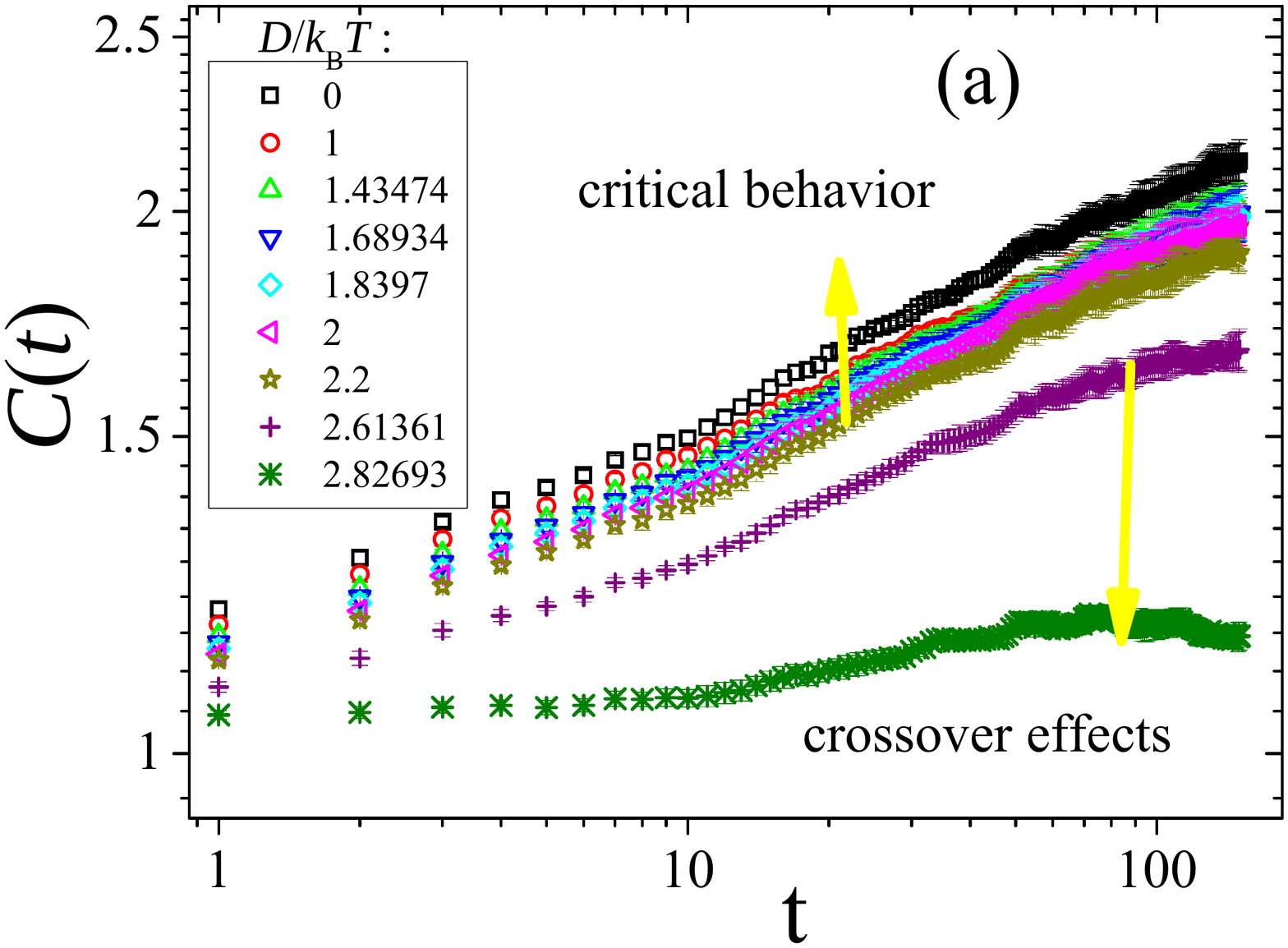} %
\includegraphics[width=1.0\columnwidth]{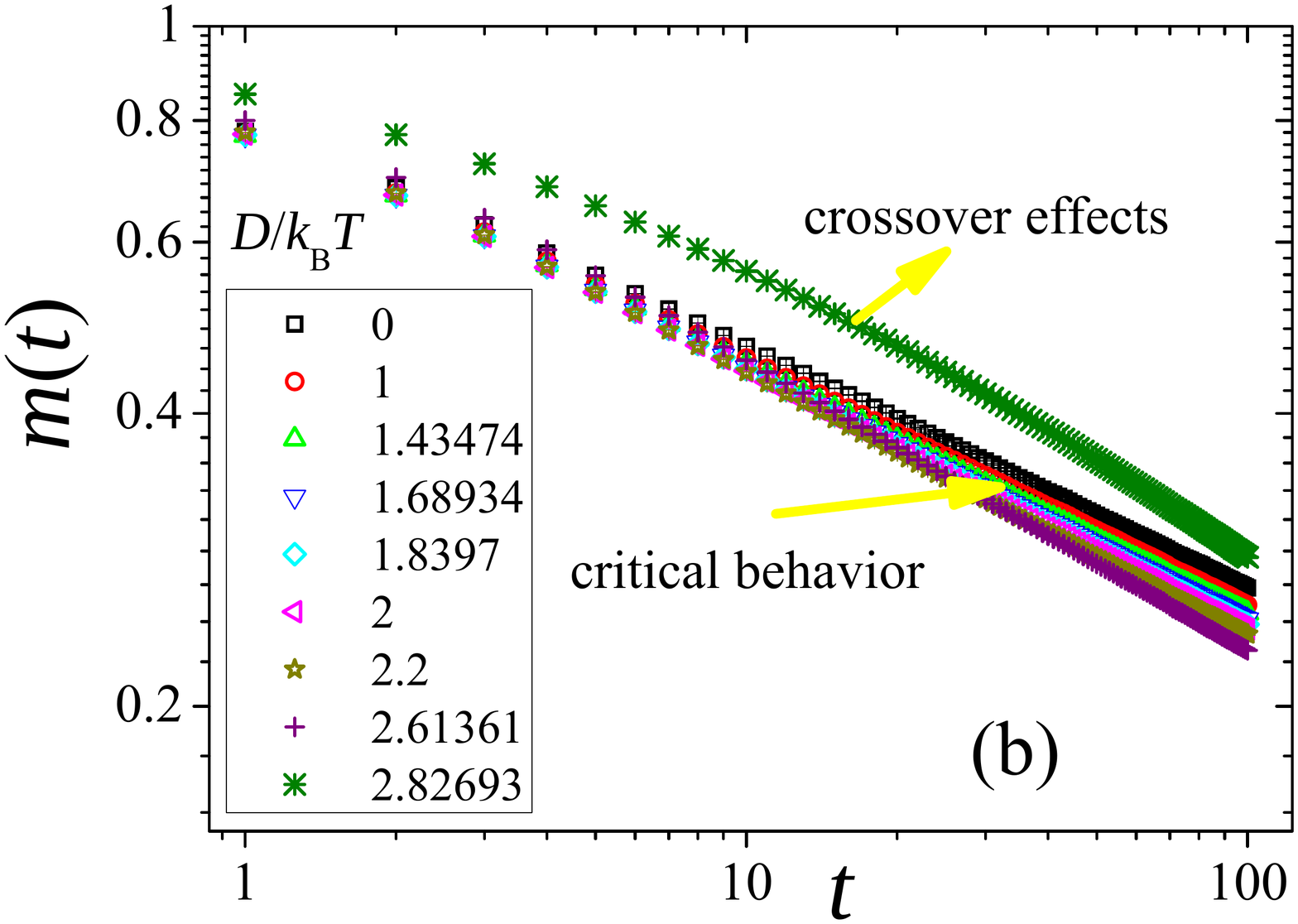} %
\includegraphics[width=1.0\columnwidth]{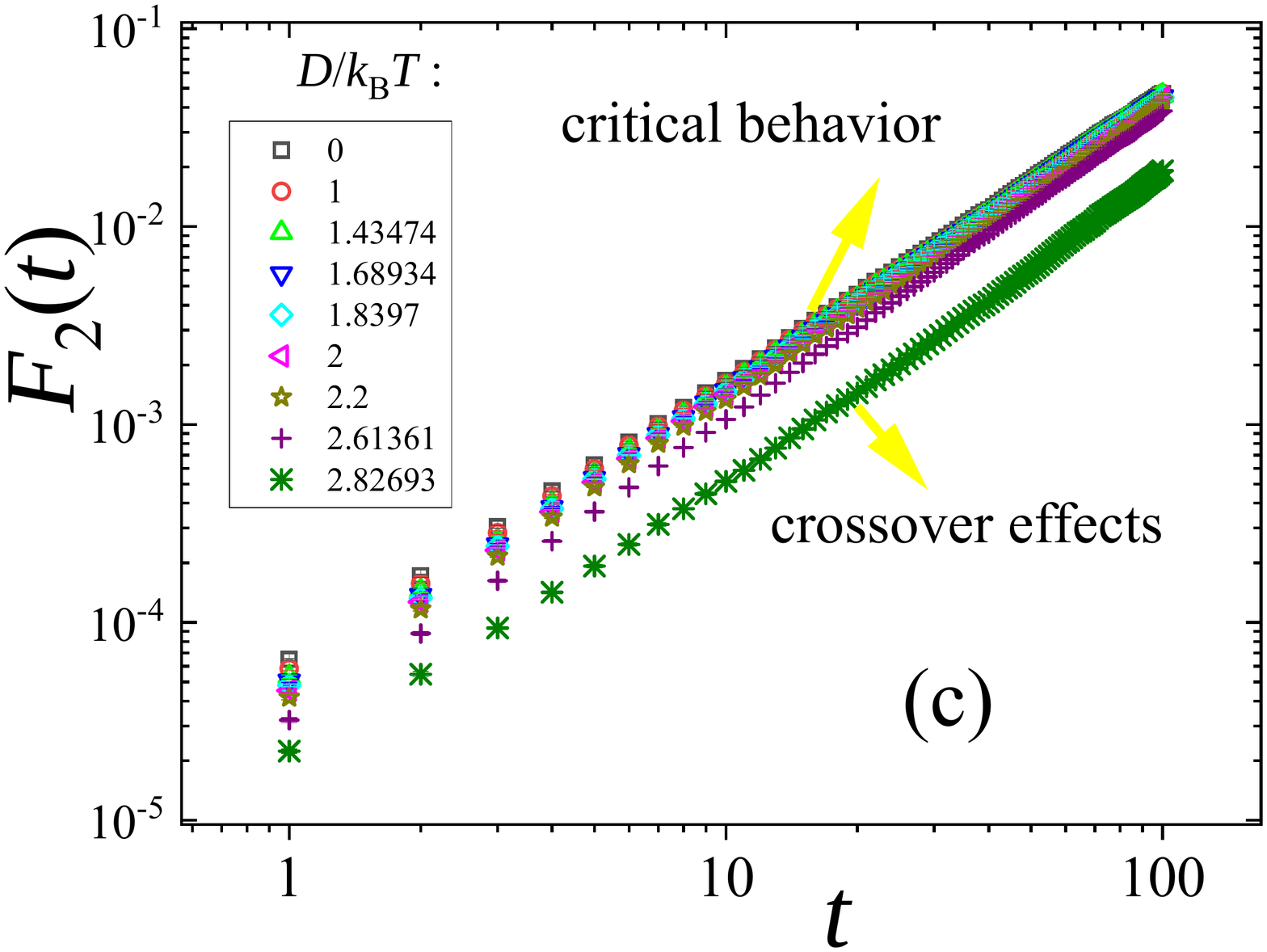}
\end{center}
\caption{(a) Time evolution of $C(t)$. (b) Time evolution of $m(t)$ for $%
m_{0}=1$. (c) $F_{2}(t)\times t$. The used points are the same ones that we
refined in Fig. \protect\ref%
{Fig:critical_line_optimization_comparison_Butera}}
\label{Fig:time_evolutions}
\end{figure}

Fig. \ref{Fig:time_evolutions} (a), (b), and (c) shows the time evolutions
of $C(t)$, $m(t)$ for $m_{0}=1$, and $F_{2}(t)$ respectively, for $D/J=0$,$\
1$,$\ 1.43474$, $1.68934$, $1.8397$,$\ 2$,$\ 2.2$,$\ 2.61361$, and $2.82693$%
, corresponding to the critical temperatures given respectively by: $%
k_{B}T/J=$ $3.19622$, $2.877369$, $2.7$, $2.57914$, $2.5$, $2.407314$, $%
2.275495$, $1.9$, and $1.5$. It is interesting to observe those power laws
present changes when they approach tricritical point: $D/J=2.84479\ \
k_{B}T/J=1.4182$. These crossover effects, visually observed in these plots,
can also be numerically checked.

%TCIMACRO{\TeXButton{B}{\begin{table}[tbp] \centering}}%
%BeginExpansion
\begin{table*}[tbp] \centering%
%EndExpansion
\begin{tabular}{l|lllllll|ll}
\hline\hline
$\frac{D}{J}$ & 0 & 1 & 1.43474 & 1.68934 & 1.8397 & 2.0 & 2.2 & 2.61361 & 
2.82693 \\ \hline\hline
$\lambda $ & 0.2492(14) & 0.2536(11) & 0.2565(15) & 0.2585(13) & 0.26164(74)
& 0.2651(13) & 0.26873(82) & \textbf{0.2984(10)} & \textbf{0.3035(64)} \\ 
$z$ & 2.068(14) & 2.051(14) & 2.037(13) & 2.022(12) & 2.029(18) & 2.005(26)
& 2.006(19) & \textbf{1.928(12)} & \textbf{1.8865(66)} \\ 
$\theta $ & 0.111(16) & 0.091(16) & 0.114(11) & 0.112(14) & 0.112(13) & 
0.110(15) & 0.1080(68) & \textbf{0.081(15)} & \textbf{0.003(11)} \\ 
$\beta /\nu $ & 0.5153(45) & 0.5201(46) & 0.5225(45) & 0.5227(39) & 
0.5309(49) & 0.5315(74) & 0.5391(54) & \textbf{0.5753(41)} & \textbf{%
0.573(23)} \\ \hline\hline
\end{tabular}%
\caption{Exponents of the Blume Capel model along the critical line obtained
with TDMC simulations. The bold results highlight the crossover effects}%
\label{Table:critical_exponents}%
%TCIMACRO{\TeXButton{E}{\end{table}}}%
%BeginExpansion
\end{table*}%
%EndExpansion

To do that, let us check the exponents shown in table \ref%
{Table:critical_exponents} analyzing their universality. One has only values
for $D=0$ in the literature. For example, $\theta $ calculated by Jaster et
al. \cite{Jaster} by directly analyzing the initial slip of the
magnetization $m(t)=m_{0}t^{\theta }$, performing $m_{0}\rightarrow 0$
yields $\theta =0.108(2)$ which is in agreement with our estimate.
Similarly, these same authors obtained $z=2.042(6)$ that agrees with our
estimate with two uncertainty bars. Finally, these authors obtained $\beta
/\nu =0.517(2)$ which agrees with our estimates. It is essential to mention
that we obtained larger error bars, considering five different bins,
corresponding to five different exponents, that, when averaged, yield our
final estimate with respective uncertainty. It is important to mention that
if we consider a unique time series with uncertainties of the points and
only then, calculating an exponent whose uncertainty comes from the linear
fit, we obtain smaller error bars. Here we opted by using the more
conservative method (first) with larger error bars.

We observe a slight variation of the exponents $\lambda $, $z$, $\theta $
and $\beta /\nu $ up to 2.2. However, after this value, the crossover
effects are pretty sensitive for $\frac{D}{J}=$ 2.61361 and 2.82693, which
corroborates which one visually observed in Fig. \ref{Fig:time_evolutions}.
Thus one can conclude that the law described by Eq. \ref{Eq:power_law_1} is
suitable to describe critical points of the Blume Capel in three dimensions
and crossover effects with $\theta >0$. It would be suggestive to think that
for the tricritical point as in two dimensions, we should find similar law
to Eq. \ref{Eq:power_law_1} but with $\theta <0$. However, it does not occur
for tricritical points from three-dimensional systems. Janssen and Oerding 
\cite{Janssen1994} demonstrated that such a problem demands logarithmic
corrections to explain the relaxation dynamics. Nevertheless, the question
is: can we observe this behavior via time-dependent MC simulations? The
answer is positive, and we will show how to perform it, which is the most
important point of this paper, and it requires a suitable numerical
exploration.

\begin{figure}[tbp]
\begin{center}
\includegraphics[width=1.0\columnwidth]{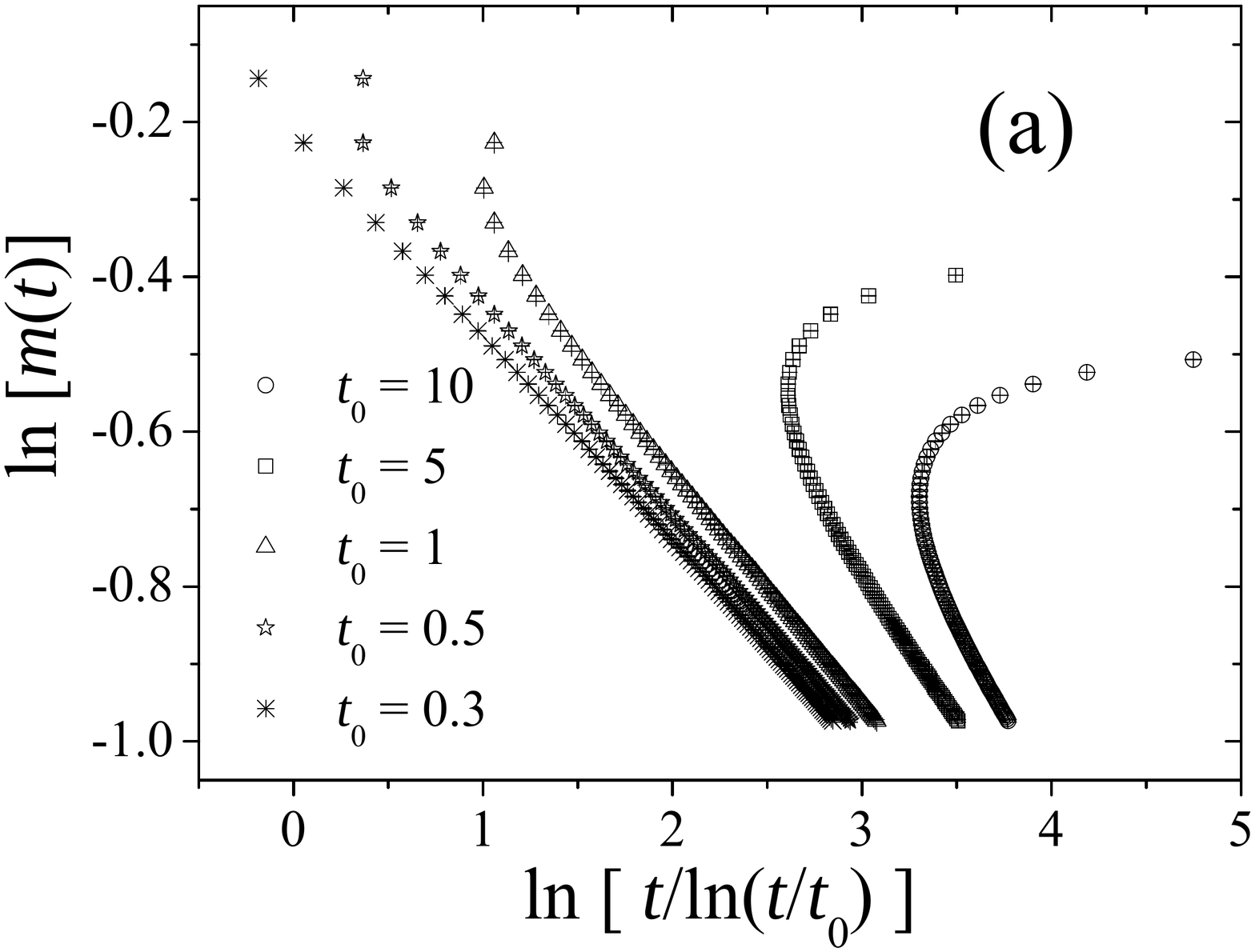} %
\includegraphics[width=1.0\columnwidth]{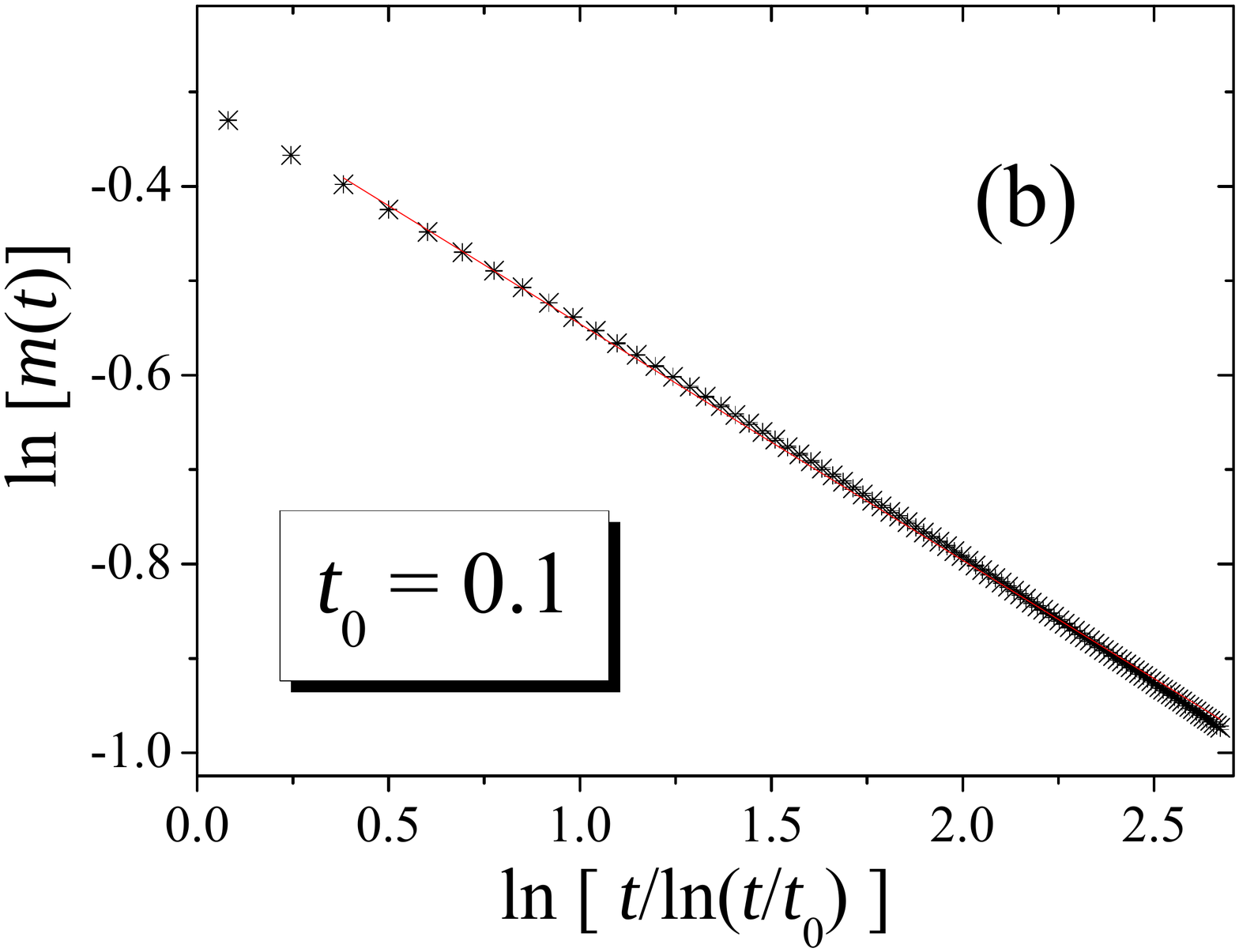}
\end{center}
\caption{(a) Time evolution of $m(t)$ at the tricritical point, considering
different values of $t_{0}$. (b) Time evolution of $m(t)$ for the particular
case $t_{0}=0.1$. }
\label{Fig:tricritical_point_decay_ferro}
\end{figure}

The results obtained by Janssen and Oerding \cite{Janssen1994}, using
methods of renormalized field theory, suggest (after some simple
manipulations) that magnetization, for three dimensions, at tricritical
point, behaves as:%
\begin{equation}
m_{3D-\text{TP}}(t)=m_{0}\frac{\ln (t/t_{0})^{-a}}{\left[ 1+\left( \frac{t}{%
\ln (t/t_{0})}\right) \ln (t/t_{0})^{-4}m_{0}^{4}\right] ^{4}}
\label{Eq:Scaling_TP}
\end{equation}%
According to this theory, $a$ is precisely given by $\frac{3}{40\pi }$. Thus
the order parameter (magnetization) must present a crossover between a pure
logarithmic behavior for short times followed by a power law with
logarithmic corrections:

\begin{equation}
m_{3D-\text{TP}}(t)=\left\{ 
\begin{array}{lll}
m_{0}\left( \ln \left( t/t_{0}\right) \right) ^{-a} & \text{for } & 
t_{0}<t<<m_{0}^{-4} \\ 
&  &  \\ 
\left( \frac{t}{\ln \left( t/t_{0}\right) }\right) ^{-\frac{1}{4}} & \text{%
for} & t>>m_{0}^{-4}\text{.}%
\end{array}%
\right.  \label{Eq:TP_law}
\end{equation}%
Here, $t_{0}$ is the microscopic time scale. Nevertheless, we perform
time-dependent MC simulations for the TP of the three-dimensional
Blume-Capel model. One starts by analyzing the relaxation from $m_{0}=1$. In
order to capture the the behavior $m_{3D-\text{TP}}(t)\sim $ $\left( \frac{t%
}{\ln \left( t/t_{0}\right) }\right) ^{-\frac{1}{4}}$. In this case, it is
interesting to change $t_{0}$ to observe the law for short times as observed
in Fig. \ref{Fig:tricritical_point_decay_ferro} (a).

The most important here is to use the correct scale. For that we performed a
plot of $\ln \left[ m(t)\right] $ versus $\ln \left[ \frac{t}{\ln \left(
t/t_{0}\right) }\right] $. We can observe that for lower $t_{0}$ values, we
observe prolonged linear behavior. The Fig. \ref%
{Fig:tricritical_point_decay_ferro} (b) shows the particular case ($%
t_{0}=0.1 $) used to measure the slope that must be 1/4 according to
prediction obtained by Janssen and Oerding. One finds $\xi =0.25034(53)$
corroborating the prediction.

For the second part, we performed simulations for small values of $m_{0}$.
However, obtaining reasonable estimates for small values of $m_{0}$ is
numerically complicated due to the fluctuations. Thus, we used $m_{0}=0.08$, 
$0.06$, $0.04$, and $0.02$. We show the time evolutions in Fig. \ref%
{Fig:tricritical_point_small_m0} (a).

\begin{figure}[tbp]
\begin{center}
\includegraphics[width=1.0\columnwidth]{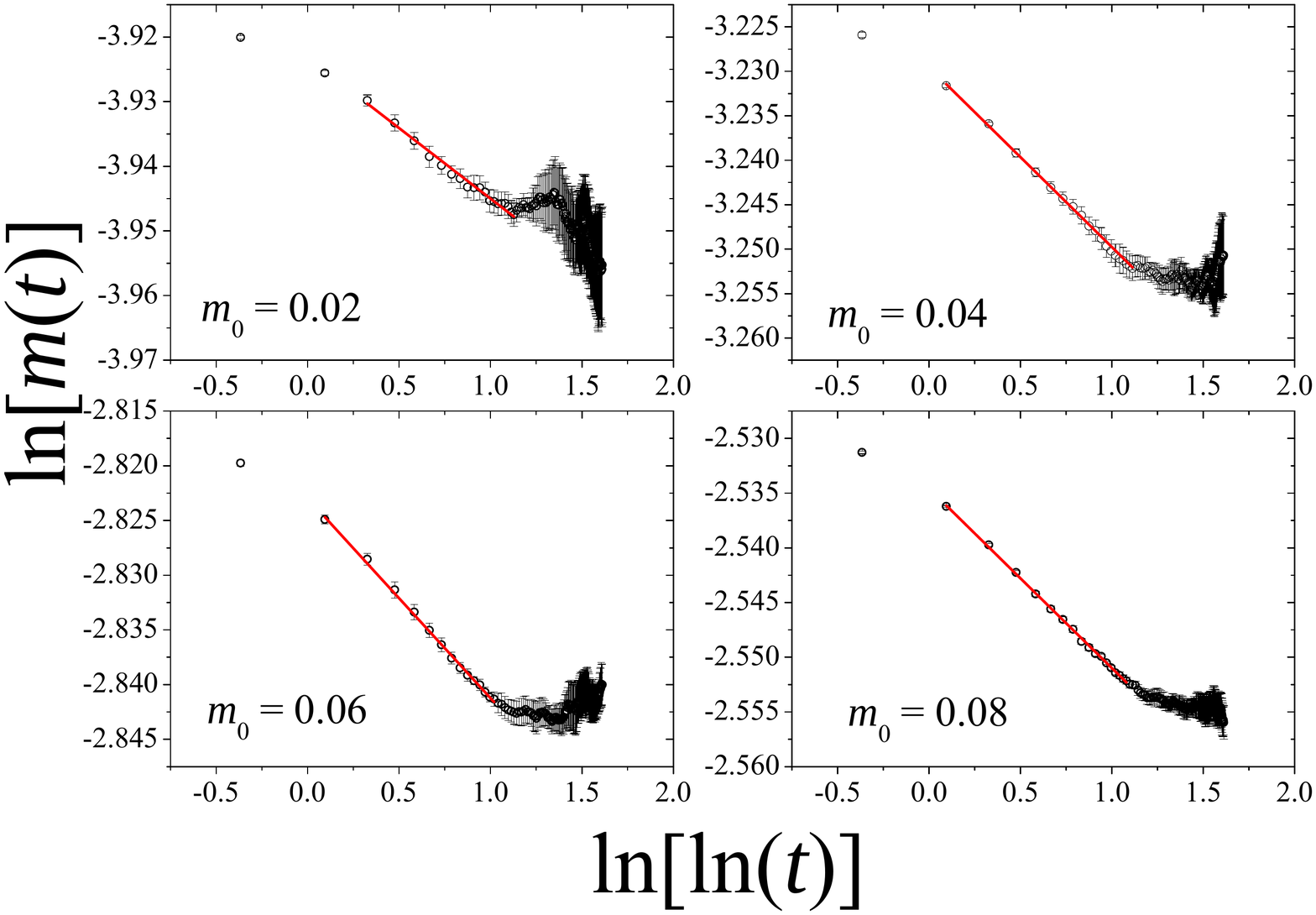} %
\includegraphics[width=1.0\columnwidth]{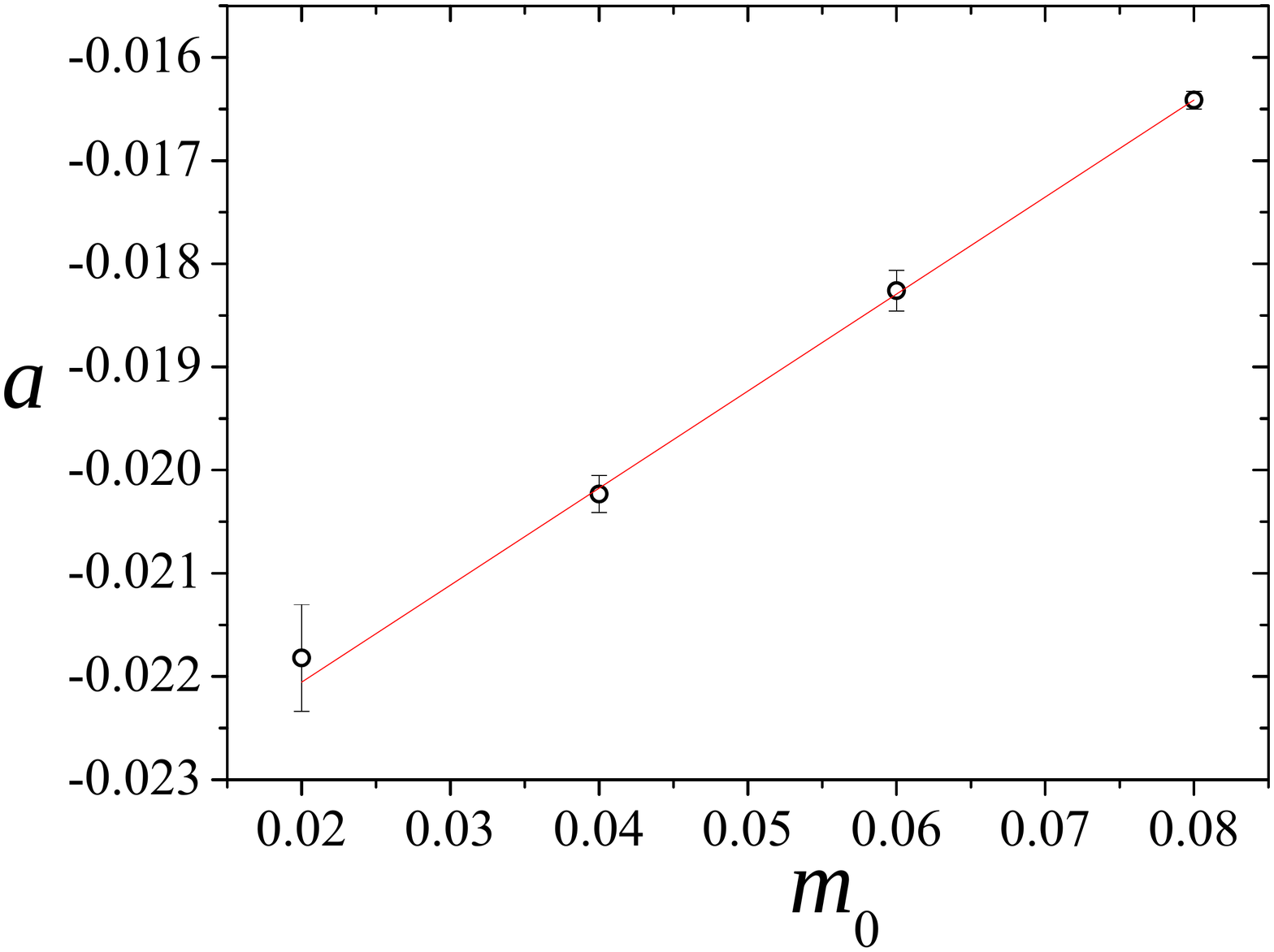}
\end{center}
\caption{(a) Time evolutions at the tricritical point for values $m_{0}=0.02$%
, $0.04$, $0.06$, and $0.08\ $ (b) Numerical extrapolation of the exponent $%
a.$}
\label{Fig:tricritical_point_small_m0}
\end{figure}

Thus we measured the slopes in the possible regions where one observed a
reasonable short duration linear behavior in the plot of $\ln (m(t))$ $%
\times $ $\ln (\ln (t))$, for different values of $m_{0}$. See the straight
lines (in red) as indicated in the figure \ref%
{Fig:tricritical_point_small_m0} (a). The slopes supposedly supply the value
of the exponent $a$ according to Eq. \ref{Eq:TP_law}. We also observe a
linear behavior of $a$ as a function of $m_{0}\ $(see Fig. \ref%
{Fig:tricritical_point_small_m0} (b) ). With this in hand, one can perform
an extrapolation for $m_{0}\rightarrow 0$. Such extrapolation yields our
estimate $a_{estimated}=0.02393(13)$, in good agreement when compared with
the theoretical prediction: $a=\frac{3}{40\pi }\approx \allowbreak
0.023\,873 $.

It is also interesting to use the decay $m(t)\sim \left( \frac{t}{\ln \left(
t/t_{0}\right) }\right) ^{-\frac{1}{4}}$ expected from ordered initial
states ($m_{0}=1$) to obtain the tricritical parameters. In this case, we
must change the coefficient of determination to: 
\begin{equation}
r=\frac{\sum\limits_{t=t_{\min }}^{t_{\max }}\left( \overline{\ln m}-a-b\ln
\left( \frac{t}{\ln \left( t/t_{0}\right) }\right) \right) ^{2}}{%
\sum\limits_{t=t_{\min }}^{t_{\max }}(\overline{\ln m}-\ln m(t))^{2}}\text{.}
\label{Eq:determination_tricritical}
\end{equation}%
Based on this amount, obtained in \cite{Deserno}, we performed two
experiments: one fixed \ $D/J=2.84479$ by varying $k_{B}T/J$, and
alternatively by fixing $k_{B}T/J=1.4182$, one varies $D/J$. 
\begin{figure}[tbp]
\begin{center}
\includegraphics[width=1.0\columnwidth]{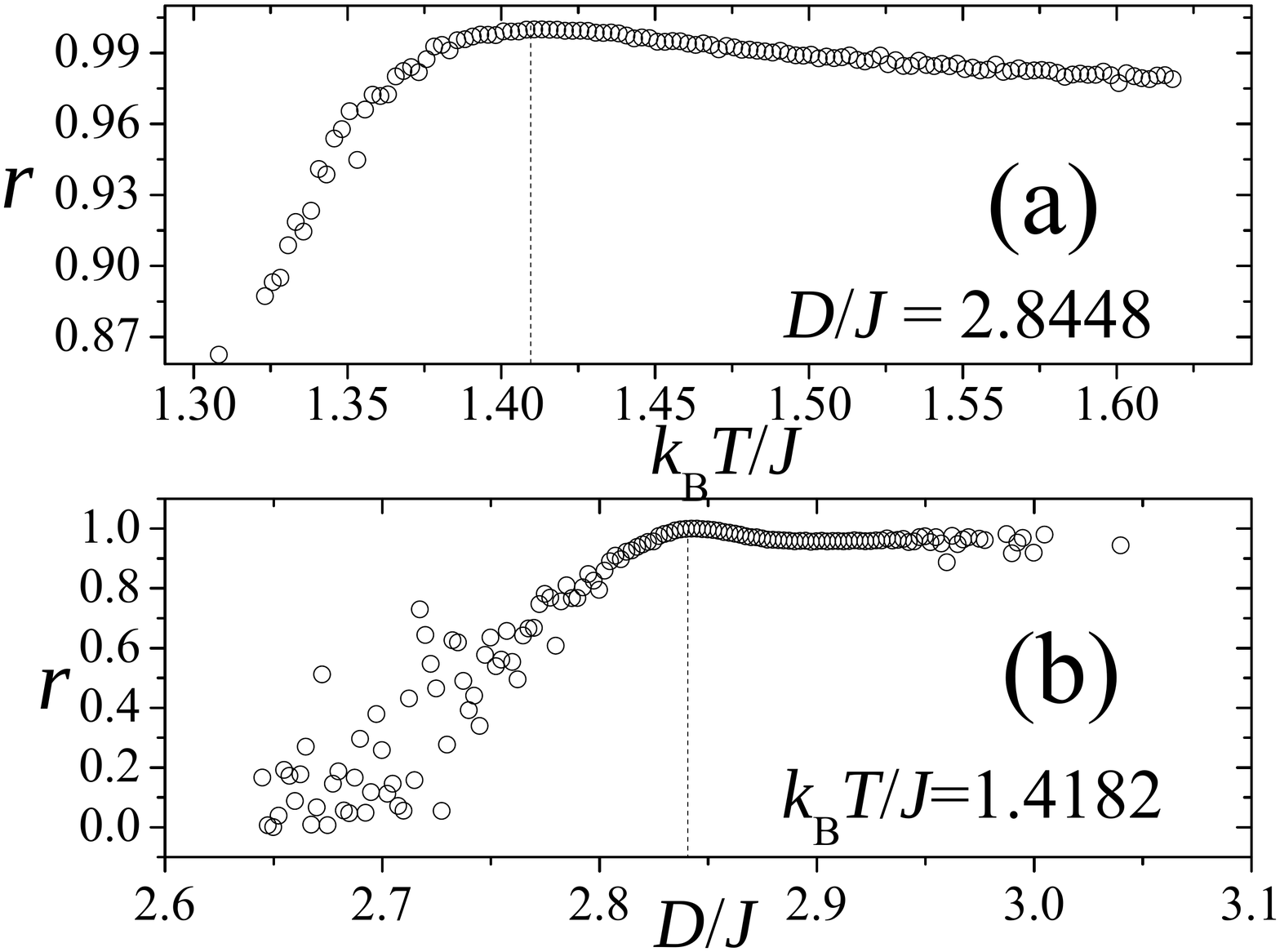}
\end{center}
\caption{(a) Coefficient of determination as a function of $k_{B}T/J$
considering $D/J$ fixed in $2.84479$. (b) Coefficient of determination as a
function of $D/J$ fixing $k_{B}T/J=1.4182$. }
\label{Fig:tricritical_localization}
\end{figure}
Fig. \ref{Fig:tricritical_localization} (a), and (b) shows both situations
respectively. The optimal values correspond to the maximal $r$,
corroborating the estimates for the TP from literature (see for example \cite%
{Deserno,Butera}), showing that our refinement method can be modified to
attend the temporal laws at TP, i.e., including the log-corrections.

It is interesting to observe that if \ the magnetization relaxes at TP as a
power-law $t^{-1/4}$ with additional logarithimic corrections, starting from 
$m_{0}=1$, the system seems to predict what happens in the mean-field
regime, since that in a recent work, we considered that evolution of
magnetization in such regime follows the differential equation \cite%
{Silva2021}: 
\begin{equation*}
\frac{dm}{dt}=-m+\frac{2e^{-\beta D}\sinh (\beta Jzm)}{2e^{-\beta D}\cosh
(\beta Jzm)+1}\text{.}
\end{equation*}%
From a very simplified point of view, such an equation leads to a crossover
between a power-law $m(t)\sim t^{-1/2}$ at the CL to a power-law $m(t)$ $%
\sim t^{-1/4}$ at the TP. Thus, the \textquotedblleft
trace\textquotedblright\ of this exponent 1/4, which must occur for $d\geq 4$
\cite{Pleimling,Lawrie}, would already appear in three dimensions but with
logarithmic corrections.

In summary, this paper verifies the theoretical predictions which suggest
log-corrections for the TP \cite{Janssen1994}. We also obtained the critical
exponents for the CL in three dimensions. One observes the crossover effects
using time-dependent MC simulations, considering the time evolution of
different amounts as time-correlation, the ratio with different initial
conditions, and the direct time evolution of magnetization. Our predictions
suggest that mean-field behavior has some brief similarities with
three-dimensional results suggested by a recent mean-field study developed
in \cite{Silva2021}. \qquad \qquad

\textbf{Acknowledgements} R. da Silva thanks CNPq for financial support
under grant numbers 311236/2018-9, 424052/2018-0, and 408163/2018-6.

\end{document}